\documentclass[preprint,amssymb]{revtex4-1}
\usepackage{graphicx}
\usepackage{natbib}

\begin{document}

\title{Effects of polymer additives on Rayleigh-Taylor turbulence}

\author{G. Boffetta$^{1,2}$, A. Mazzino$^{3}$ and S. Musacchio$^{4}$}
\affiliation{$^1$Dipartimento di Fisica Generale and INFN, 
Universit\`a di Torino, via P.Giuria 1, 10125 Torino (Italy) \\
$^2$ISAC-CNR, corso Fiume 4, 10133 Torino (Italy) \\
$^3$Dipartimento di Fisica, Universit\`a di Genova, INFN and CNISM,
via Dodecaneso 33, 16146 Genova (Italy) \\
$^{4}$ CNRS, Lab. J.A. Dieudonn\'e UMR 6621,
Parc Valrose, 06108 Nice (France)}

\begin{abstract}
The role of polymers additives on the turbulent convective flow of a
Rayleigh--Taylor system is investigated by means of direct numerical
simulations (DNS) of Oldroyd-B viscoelastic model. The dynamics of polymers
elongation follow adiabatically the self-similar evolution of the turbulent
mixing layer, and shows the appearance of a strong feedback on the flow which
originate a cut off for polymer elongations. The viscoelastic effects on the
mixing properties of the flow are twofold. Mixing is appreciably enhanced at
large scales (the mixing layer growth-rate is larger than that of the purely
Newtonian case) and depleted at small scales (thermal plumes are more coherent
with respect to the Newtonian case). The observed speed up of the thermal
plumes, together with an increase of the correlations between temperature field
and vertical velocity, contributes to a significant {\it enhancement of heat
transport}. 
Our findings are consistent with a scenario of {\it drag reduction}
between falling and rising plumes induced by polymers, and provide further
evidence of the occurrence of drag reduction in absence of boundary layers.
A weakly non-linear model proposed by Fermi for the growth of the mixing layer 
is reported in the Appendix.
\end{abstract}

\maketitle

\section{Introduction}
Polymer additives have dramatic effects on the dynamics of turbulent flows,
the most important being the reduction of turbulent drag
up to 80 $\%$ when few parts per million of long-chain
polymers are added to water \cite{toms49}.
The paramount relevance of this phenomenon motivated the strong 
efforts of researchers aimed to achieve a better understanding 
of the basic mechanisms of polymer drag reduction.
The natural framework of drag-reduction studies is the case of pipe flow, 
or channel flow. Within this context the reduction of the frictional 
drag against material wall originated by the addition of polymers, 
manifests as an increase of the mean flow across the pipe or 
channel at given pressure drop. 

Recent studies (see, e.g., \cite{bcm05,BBCMPV07}) showed that a drag-reduction
phenomenon may also occurs in the absence of physical walls.  In this case the
drag which is reduced is not the frictional drag against the boundaries of the
flow, but the turbulent drag of the bulk flow itself. In particular, in the
case of homogeneous isotropic turbulence, it has been observed a reduction of
the rate of energy dissipation at fixed kinetic energy associated with a
reduction of velocity fluctuations at small
scales\cite{bdgp_pre03,dcbp_jfm05,bbbcm_epl06}.  
In turbulent systems with a non-vanishing local mean flow (e.g the Kolmogorov
flow), it has been shown that polymers causes a reduction of the Reynolds
stresses which results in an increased intensity of the mean velocity 
profile~\cite{bcm05}.
This phenomenon, which occurs in absence of boundaries, 
is remarkably similar to increase of throughput observed in pipe or channel
flows, and suggest the existence of common features and possibly of common
physical mechanisms between the drag-reduction occurring in wall-bounded and in
bulk flows. 

In the present paper we provide further evidence of turbulent drag reduction 
in bulk flows by studying the effects of polymers additives in the
Rayleigh-Taylor (RT) setup of turbulence convection.  A previous study
\cite{bmmv_jfm10} has already shown that polymers affect the early stage
(linear phase) of the RT instability which occurs at the interface between two
unstably stratified fluids. Here, we show that polymers also induce strong
modifications in the dynamics of the turbulent mixing layer, which develops in
the late stage of the mixing process. 
In particular we study how polymers are able to affect the process of 
turbulent heat transfer with a mechanism which is probably more general 
than the particular case studied here.

Preliminary results have been presented in \cite{bmmv_prl10} and are
briefly reported here for completeness. We provide here new results 
supporting our interpretation of the mechanism at the basis of the 
observed effects together with results on polymer statistics and
small scale turbulence.

The remaining of this paper is organized as follow.  In Section~\ref{sez:0} we
introduce the viscoelastic Rayleigh--Taylor problem and give some details on
the numerical strategy we exploited to study polymer dynamics.  In
Section~\ref{sez:1} we analyze the statistics of polymer elongations.  In
Section~\ref{sez:2} we show the effects of polymers on the turbulent mixing.
In Section~\ref{sez:3} we discuss the drag reduction phenomenon in the
viscoelastic RT.  In Section~\ref{sez:4} we study the effects induced by
polymers on the heat transport. 
Conclusions are devoted to a short discussion on the possibility to observe
the described effects in the laboratory. 
Finally, in the Appendix we briefly describe the model for the growth of 
the mixing layer proposed by Fermi.

\section{The viscoelastic Rayleigh--Taylor model}
\label{sez:0}
We will focus our attention on the miscible case of the RT system at low Atwood
number and Prandtl number one.  Within the Boussinesq approximation, generalized
to a viscoelastic fluid using the standard Oldroyd-B model\cite{bhac87}, the
equations for the dynamics of the velocity field ${\bf u}$ coupled to the
temperature field $T({\bf x},t)$ (which is proportional to the density via the
thermal expansion coefficient $\beta$ 
as $\rho=\rho_0 [1-\beta (T-T_0)]$, $\rho_0$ and $T_0$ are reference
values) and the positive symmetric conformation tensor of polymers
$\sigma_{ij}({\bf x},t)$ read: 
\begin{eqnarray}
\partial_t {\bf u} + {\bf u} \cdot {\bf \nabla} {\bf u} &=&
- {\bf \nabla} p + \nu \nabla^2 {\bf u} - \beta {\bf g} T + 
{2 \nu \eta \over \tau_p}
{\bf \nabla} \cdot \sigma \nonumber  \\
\partial_t T + {\bf u} \cdot {\bf \nabla} T &=& \kappa \nabla^2 T 
\label{eq:1} \\
\partial_t \sigma + {\bf u} \cdot {\bf \nabla} \sigma &=&
({\bf \nabla} {\bf u})^{T} \cdot \sigma + \sigma \cdot 
({\bf \nabla} {\bf u}) - {2 \over \tau_p} (\sigma - \mathbb{I}) + 
\kappa_p \nabla^2 \sigma \nonumber
\end{eqnarray}
together with the incompressibility condition 
${\bf \nabla} \cdot {\bf u} = 0$.
In (\ref{eq:1}) ${\bf g}=(0,0,-g)$ is gravity acceleration, 
$\nu$ is the kinematic viscosity, 
$\kappa$ is the thermal diffusivity,
$\eta$ is the zero-shear polymer contribution to total viscosity
$\nu_T=\nu(1+\eta)$
(proportional to polymers concentration)
and $\tau_p$ is the (longest) polymer relaxation time,
{\it i.e.} the Zimm relaxation time for a linear chain 
$\tau_p=\nu R_0^3/(\rho k_B T)$ with $k_B$ Boltzmann constant
and $R_0$ the radius of gyration \cite{bhac87}. 
The diffusive term $\kappa_p \nabla^2 \sigma$ is added to
prevent numerical instabilities \cite{sb_jnnfm95}.

The initial condition for the RT problem is an unstable temperature jump
$T({\bf x},0)=-(\theta_0/2)sgn(z)$ in a fluid at rest ${\bf u}({\bf x},0)=0$
with coiled polymers $\sigma({\bf x},0)=\mathbb{I}$.  The physical assumptions
under which the set of equations (\ref{eq:1}) is valid are of small Atwood
number $A=(1/2) \beta \theta_0$ and dilute polymers solution.  Experimentally,
density fluctuations can also be obtained by some additives (e.g., salt)
instead of temperature fluctuations: within the validity of Boussinesq
approximation, these situations are described by the same set of equations
(\ref{eq:1}).  In the following, all physical quantities are made dimensionless
using the vertical side, $L_z$, of the computational domain, the temperature
jump $\theta_0$ and the characteristic time $\tau=(L_z/A g)^{1/2}$ as
fundamental units.

Numerical simulations of equations (\ref{eq:1}) have been performed with a
parallel pseudospectral code, with $2$-nd order Runge-Kutta time scheme on a
discretized domain of $N_x \times N_y \times N_z$ grid points.  Periodic
boundary conditions in all directions are imposed.  
The initial perturbation is seeded in both cases by adding a
$10\%$ of white noise (same realization for both runs) 
to the initial temperature profile in a small
layer around the instable interface at $z=0$. 
Because of periodicity
along the vertical direction, the initial temperature profile has two
temperature jumps: an unstable interface at $z=0$ which develops in the
turbulent mixing layer and a stable interface at at $z=
\pm L_z/2$.  Numerical simulations are halted when the mixing layer is still
far from the stable interface, whose presence has no detectable influence on
the simulations (velocities there remain close to zero).  The results of the
reference Newtonian simulation (denoted by run $N$)  are compared with those of
three viscoelastic runs ($A$,$B$ and $C$) with identical parameters and
different polymer relaxation time (see Table~\ref{table1}).

\begin{table}
\begin{tabular}{c|cccccccccc} 
Run & $N_{x,y}$ & $N_z$ & $L_{x,y}$ & $L_z$ & $\theta_0$ & $\beta g$ & $\nu=\kappa$& $\kappa_p$ & $\eta$ & $\tau_p$ \\
\hline
N & $512$ & $1024$ & $2\pi$ & $4\pi$ & $1$ & $0.5$ & $3\cdot 10^{-4}$ & $-$ & $-$ & $-$ \\
A & $512$ & $1024$ & $2\pi$ & $4\pi$ & $1$ & $0.5$ & $3\cdot 10^{-4}$ & $10^{-3}$ & $0.2$ & $1$ \\
B & $512$ & $1024$ & $2\pi$ & $4\pi$ & $1$ & $0.5$ & $3\cdot 10^{-4}$ & $10^{-3}$ & $0.2$ & $2$ \\
C & $512$ & $1024$ & $2\pi$ & $4\pi$ & $1$ & $0.5$ & $3\cdot 10^{-4}$ & $10^{-3}$ & $0.2$ & $10$ \\
\end{tabular}
\caption{Parameters of the simulations} 
\label{table1}
\end{table}

\section{Statistics of polymer elongations}
\label{sez:1}
Before presenting the results of our numerics, let us discuss the theoretical
behavior expected for polymers statistics in the ``passive case'' in which
their feedback on the flow is neglected.
Recent studies of Newtonian RT turbulence
\cite{chertkov_prl03,vc_pof09,bmmv_pre09} support the picture of a Kolmogorov
scenario, in which the buoyancy forces sustain the large scale motion, but they
are overcome at small scales by the turbulent cascade process. The accelerated
nature of the system results in an adiabatic growth of the flux of kinetic
energy in the turbulent cascade $\varepsilon \simeq (Ag)^2 t$.  As a
consequence, the Kolmogorov viscous scale $\eta \simeq \nu^{3/4} \varepsilon^{-1/4}$
and its associated time-scale $\tau_{\eta} \simeq (\nu / \varepsilon)^{1/2})$ decrease
in time as $\eta \simeq \nu^{3/4} (Ag)^{-1/2} t^{-1/4}$ 
and $\tau_{\eta} \simeq \nu^{1/2} (Ag)^{-1} \sim t^{-1/2}$ respectively.

The Weissenberg number $Wi=\tau_p/\tau_{\eta}$, which measure the relative
strength of stretching due to velocity gradients and polymer relaxation, grows
as $Wi \sim t^{1/2}$.  Therefore, even if the relaxation time of polymer
$\tau_p$ is sufficiently small to keep the polymers in the coiled state in the
initial stage of the evolution, they are expected to undergo a coil-stretch
transition as the system evolves.  The Lumley scale, defined as the scale
$\ell_L$ whose characteristic time is equal to the polymer relaxation time
$\tau_{\ell_L} \simeq \ell_L / \delta_{\ell_L} u \simeq \tau_p$ grows in time as $\ell_L
\simeq Ag \tau_p^{3/2} t^{1/2}$.  In view of the fact that the turbulent inertial
range extends from the integral scale $\mathcal{L} \simeq Ag t^2$ to the dissipative
scale $\eta \sim t^{-1/4}$ the temporal evolution of the Lumley scale
guarantees that for long times one has $\eta < \ell_L < \mathcal{L}$.

It is worth noting that in two dimension the behavior 
would be the opposite. In contrast to the three-dimensional (3D) case, 
the phenomenology of RT turbulence in 2D 
is characterized by a Bolgiano scenario, 
which originate from a scale-by-scale balance 
between buoyancy and inertial forces \cite{chertkov_prl03,cmv_prl06}. 
The resulting scaling behavior of velocity increments is 
$\delta_{\ell} u \simeq (Ag)^{2/5} \ell^{3/5} t^{-1/5}$, 
which gives for the dissipative scale 
$\eta \simeq (Ag)^{-1/4} \nu^{5/8} t^{1/8}$ and 
$\tau_{\eta} \simeq (Ag)^{-1/2} \nu^{1/4} t^{1/4}$.  
Therefore in the 2D case the Weissenberg number decreases in time as 
$Wi \sim t^{-1/4}$ and polymers will eventually recover the coiled state.
The Lumley scale decay as $\ell_L \simeq Ag \tau_p^{5/2} t^{-1/2}$, and in the
late stage of the evolution will become smaller that the dissipative
scale $\eta$. 

\begin{figure}
\includegraphics[clip=true,keepaspectratio,width=10cm]{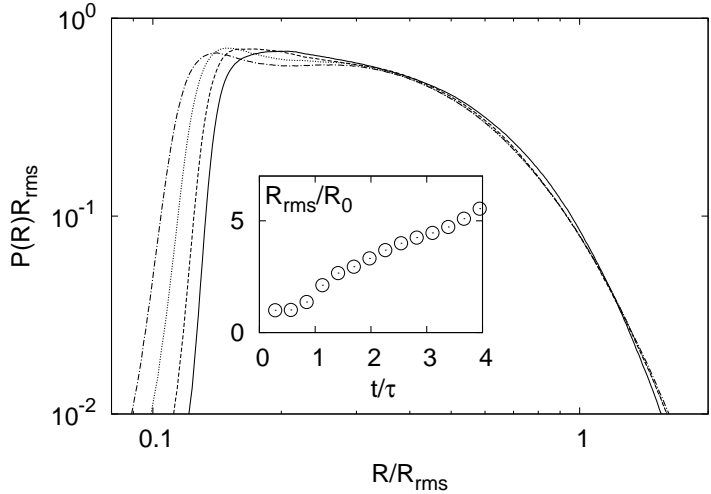}
\caption{Pdfs of polymers elongation at times $3<t/\tau<4$. 
Solid line $t/\tau=3.1$, dashed line $t/\tau=3.4$,
dotted line $t/\tau=3.7$, dash-dotted line $t/\tau=4.0$.
Inset: Rms of polymers elongation as a function of time. 
Data from run A.}
\label{fig1}
\end{figure}
\begin{figure}
\includegraphics[clip=true,keepaspectratio,width=10cm]{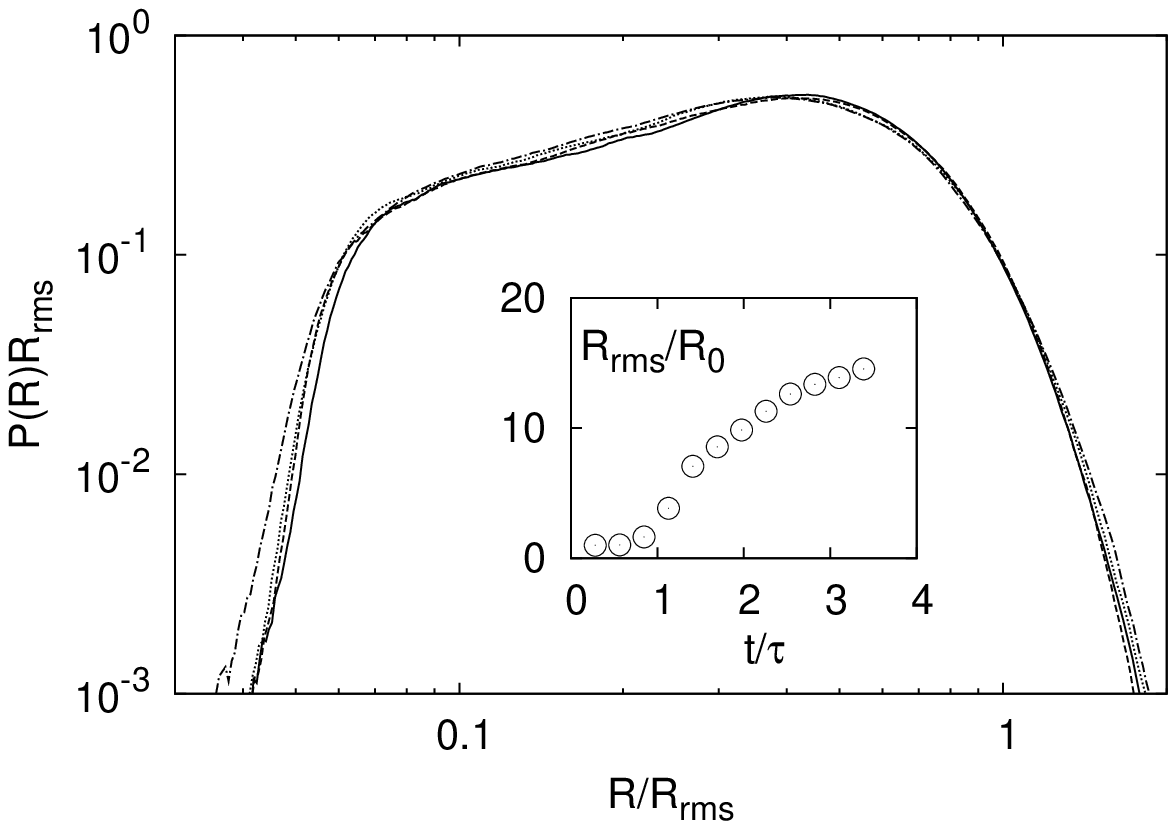}
\caption{Pdfs of polymers elongation at times $2.5<t/\tau<3.5$. 
Solid line $t/\tau=2.5$, dashed line $t/\tau=3.8$,
dotted line $t/\tau=3.1$, dash-dotted line $t/\tau=3.4$.
Inset: Rms of polymers elongation as a function of time. 
Data from run B.}
\label{fig2}
\end{figure}
\begin{figure}
\includegraphics[clip=true,keepaspectratio,width=10cm]{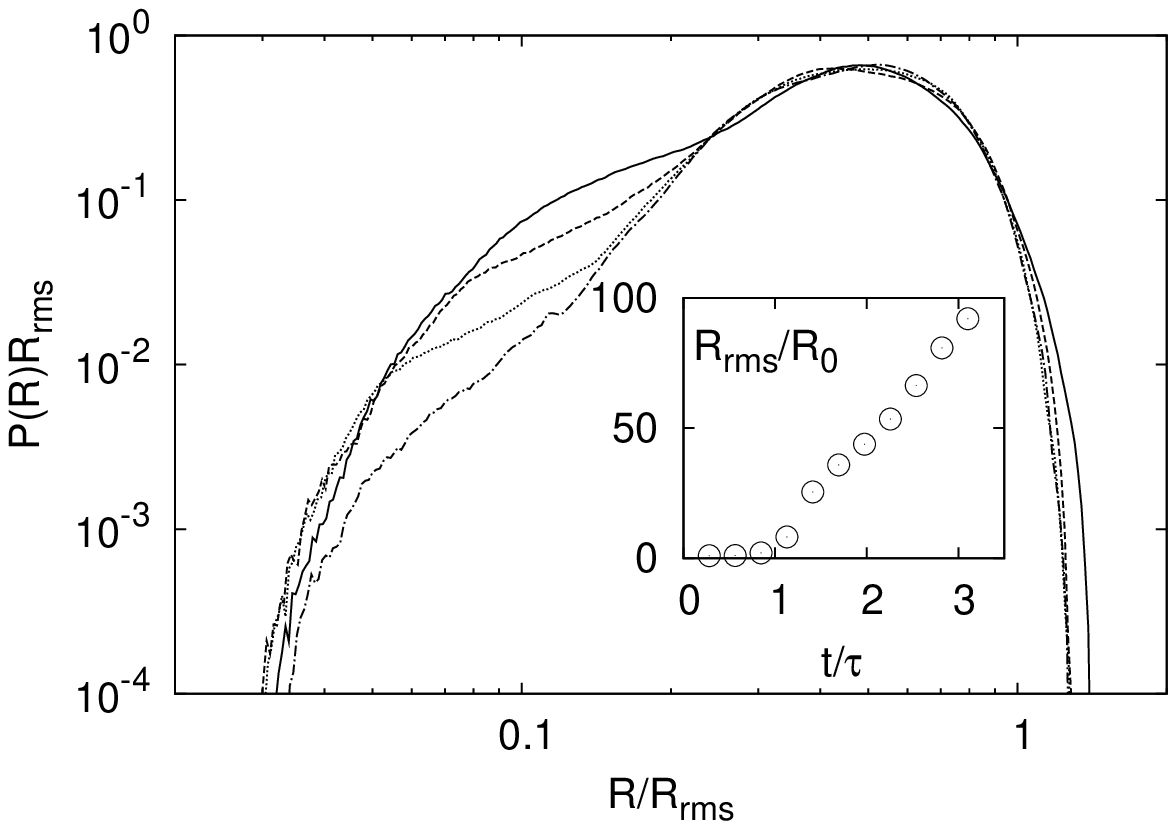}
\caption{Pdfs of polymers elongation at times $2.2<t/\tau<3.2$. 
Solid line $t/\tau=2.3$, dashed line $t/\tau=2.5$,
dotted line $t/\tau=2.8$, dash-dotted line $t/\tau=3.1$.
Inset: Rms of polymers elongation 
as a function of time. 
Data from run C.}
\label{fig3}
\end{figure}
   
Under the hypothesis that these scaling behaviors remain valid also in the
presence of polymer feedback to the flow, one may conjecture that
viscoelastic effects in 3D RT turbulence become more and more important as the
system evolves (while, as explained, in the 2D case they are expected to 
be transient, and to disappear at the late stage of the evolution).

The presence of a coil-stretch transition in the 3D RT flow 
is confirmed by the behavior of the rms polymer elongation 
$R_{rms} = \langle(tr(\sigma)/3)\rangle^{1/2} R_0$ 
measured in our simulations 
(see insets of Figs.~\ref{fig1},~\ref{fig2},~\ref{fig3}).
In the initial stage of the evolution the velocity gradients 
are too weak to significantly stretch the polymers, and $R_{rms} \sim R_0$. 
At time $t\simeq \tau $ a transition occurs, and polymers start to elongate. 
After a transient exponential growth, a regime characterized 
by a linear growth $R_{rms} \sim t$ sets in, which is consistent
with the growth of elastic energy discussed in Section~\ref{sez:3}.
The pdf of elongations in this stage of the evolution are not 
stationary, 
but their right tail collapse once rescaled with $R_{rms}$ 
(see Figs.~\ref{fig1},~\ref{fig2},~\ref{fig3}). 
Oldroyd-B model allows a priori for infinite elongations, 
but we observe an exponential cutoff for the right tail of the pdfs, 
which is a genuine viscoelastic effect: 
polymer feedback is able to reduce the stretching efficiency of the flow.
These observations lead to the  conclusion that polymers dynamics follows 
adiabatically the accelerated growth of the flow and 
generates a strong feedback which manifests in the appearance of a cutoff 
for their elongations. 

\section{Effects of polymers on mixing properties}
\label{sez:2}

The evolution of the turbulent mixing layer is strongly affected by polymer
additives.  For a Newtonian flow, because of the constant acceleration provided
by the gravity force, one expects the width $h(t)$ of the mixing layer to
grow as $h(t)= \alpha A g t^2$, where $\alpha$ is a dimensionless parameter to
be determined empirically \cite{ra_pof04,dimonte_etal_pof04,krbghca_pnas07}.
Several definitions of $h(t)$ have been proposed, based on either local or
global properties of the mean temperature profile $\overline{T}(z,t)$ (the
overbar indicates average over the horizontal directions)
\cite{as_pof90,dly_jfm99,cc_natphys06,vc_pof09}.  Here, we adopt the simplest
measure $h_r$ based on the threshold value of $z$ at which $\overline{T}(z,t)$
reaches a fraction $r$ of the maximum value i.e. 
$\overline{T}(\pm h_r(t)/2,t)= \mp r \theta_0/2$.

\begin{figure}
\includegraphics[clip=true,keepaspectratio,width=10cm]{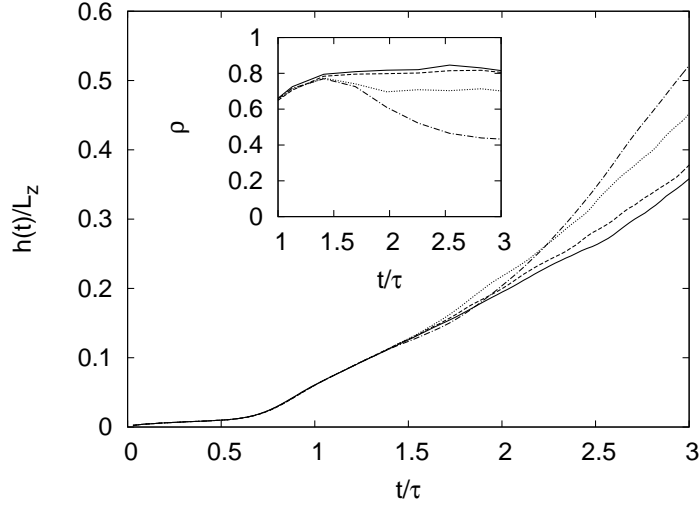}
\caption{
Temporal evolution of the mixing layer width $h(t)$. 
Inset: fraction of mixed fluid within the mixing layer
$\rho=1/(L_x L_y L_z) \int d^3 x \theta[(r \theta_0/2)^2-T^2]$.
Newtonian flow: solid line.  Viscoelastic flows: 
dashed line (Run A), dotted line (Run B), dash-dotted line (Run C). 
}
\label{fig4}
\end{figure}
\begin{figure}
\includegraphics[clip=true,keepaspectratio,width=10cm]{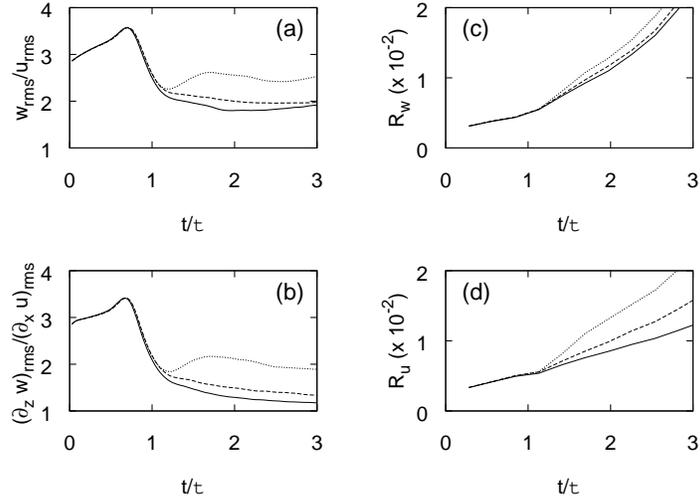}
\caption{
Time evolution of the ratio $w_{rms}/u_{rms}$ (panel a) 
and velocity gradients (panel b). 
Correlation length of vertical ($R_w$, panel c) 
and horizontal ($R_u$, panel d) velocity components. 
Newtonian flow: solid line. 
Viscoelastic flows: dashed line (Run A), dotted line (Run B). 
}
\label{fig5}
\end{figure}

In the viscoelastic solution the growth of the mixing layer is faster than in
the Newtonian case (see Fig.~\ref{fig4}), and the acceleration effect is
stronger for polymers with longer relaxation times.  On a coarse scale this
effects produces a mixing enhancement.  On the other hand, the viscoelastic
fluid is less uniformly mixed within the mixing layer itself.  In the
Newtonian case the volume of the region where $|T({\bf x},t)| < r \theta_0/2$
is roughly $80\%$ of the  volume of the mixing layer at the same time.
Conversely, the fraction of mixed fluid within the mixing layer reduces up to
$50\%$ for the viscoelastic case (see inset of Fig.~\ref{fig4}).  These
results indicate that the effects of polymers on the mixing efficiency is
twofold. At large scale they enhance the mixing by accelerating the growth of
the mixing layer and at small scale they reduce the mixing efficiency of the
turbulent flow.

These effects are accompanied by an increase of the anisotropy of the flow.  In
Figure~\ref{fig5} we show the ratio between rms of vertical ($w_rms$) and
horizontal velocities ($u_rms$) and velocity gradients.  The velocity ratio,
which is around $1.8$ for the Newtonian case \cite{bmmv_pre09}, becomes larger
than $2.5$ for the viscoelastic run. This phenomenon is associated with the
enhancement of the vertical velocity of the mixing layer.  The reduction of
small-scale mixing efficiency results in the persistence of anisotropy also at
small scales (i.e. in the velocity gradients), at variance with the Newtonian
case in which it is almost absent.  The viscoelastic flow is therefore
characterized by the presence of faster and larger plumes than those
characterizing  the Newtonian case.  The increased coherence of thermal plumes
can be quantified in terms of the enhancement of the velocity correlation
length (here defined as the half width of the velocity correlation function
\cite{vc_pof09}), of both horizontal and vertical velocity components (see
Fig.~\ref{fig5}).

\section{Interpretation in terms of drag reduction}
\label{sez:3}
\begin{figure}
\includegraphics[clip=true,keepaspectratio,width=10cm]{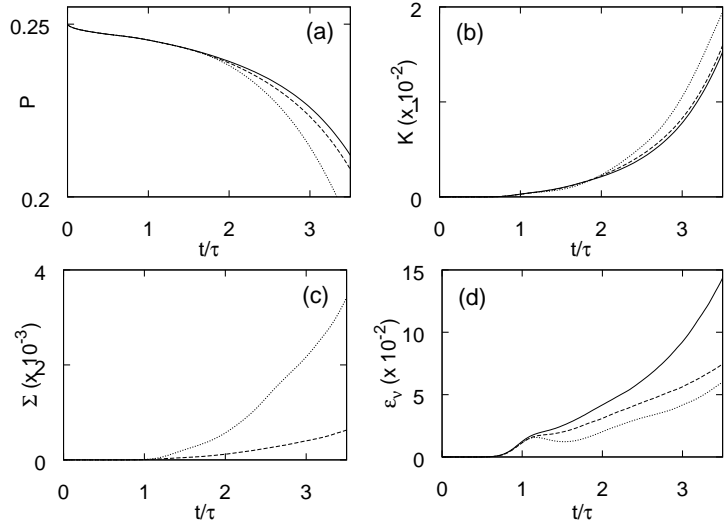}
\caption{
Temporal evolution of the potential energy $P$ (panel a), 
kinetic energy $K$ panel (b), elastic energy $\Sigma$ (panel c)
and viscous energy dissipation (panel d). 
Newtonian flow: solid line. 
Viscoelastic flows: dashed line (Run A), dotted line (Run B). 
}
\label{fig6}
\end{figure}

The energy balance of the viscoelastic RT system differs from the Newtonian 
case
because of the elastic contribution to the energy and dissipation.  The energy
can be written as the sum of potential, kinetic and elastic contributions:
\begin{equation}
E=P+K+\Sigma= 
-\beta g \langle z T \rangle 
+ \frac{1}{2} \langle u^2 \rangle 
+ \frac{\nu \eta}{\tau_p} \left[ \langle tr \sigma \rangle -3 \right]
\label{eq:2}
\end{equation}
and the energy balance reads: 
\begin{equation}
{d E \over dt} = - \varepsilon_{\nu} - \varepsilon_{\Sigma}
\label{eq:3}
\end{equation}
where 
$\varepsilon_{\nu}=\nu \langle (\partial_{\alpha} u_{\beta})^2 \rangle$
is the viscous dissipation and the last term represents elastic dissipation
$\varepsilon_{\Sigma} = 2 \Sigma / \tau_p$.
The evolution of the system is sustained by the consumption of 
potential energy, which provides a power source  
$-{d P \over dt} = \beta g \langle w T \rangle$ 
(where $w$ is the vertical velocity component).
It is worth noting that the rate of energy injection is not determined a
priori. Indeed, it is the dynamics itself which determines the rate of
conversion of potential energy into kinetic and elastic energy.  Our numerics
reveals that polymers accelerate this process (see Fig.~\ref{fig6}), and that
kinetic energy for viscoelastic runs is larger than that of 
the Newtonian case (of about $40 \%$ at $t=3.5 \tau$).  
We remark that the faster growth
of kinetic energy is not a straightforward consequence of the speed-up of
potential energy consumption, due to the accelerated growth of mixing layer.
Part of the potential energy is indeed converted
into elastic energy and finally dissipated by polymers relaxation 
to equilibrium.

\begin{figure}
\includegraphics[clip=true,keepaspectratio,width=10cm]{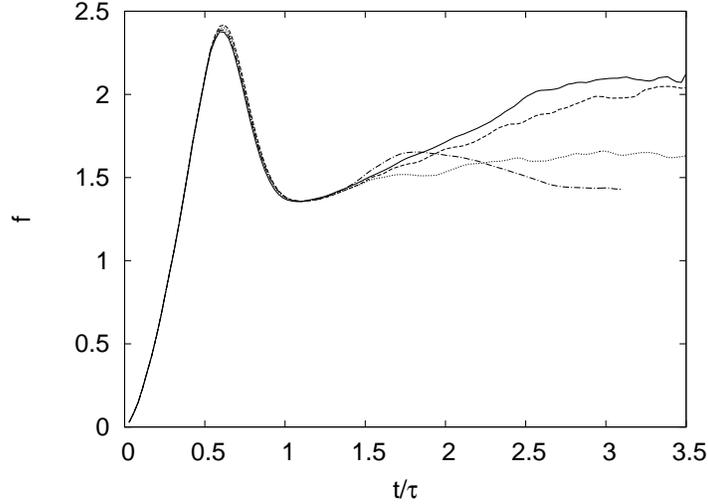}
\caption{
Time evolution of the drag reduction factor $f$. 
Newtonian flow: solid line.  
Viscoelastic flows: dashed line (Run A), dotted line (Run B), dash-dotted line (Run C).
}
\label{fig7}
\end{figure}
\begin{figure}
\includegraphics[clip=true,keepaspectratio,width=10cm]{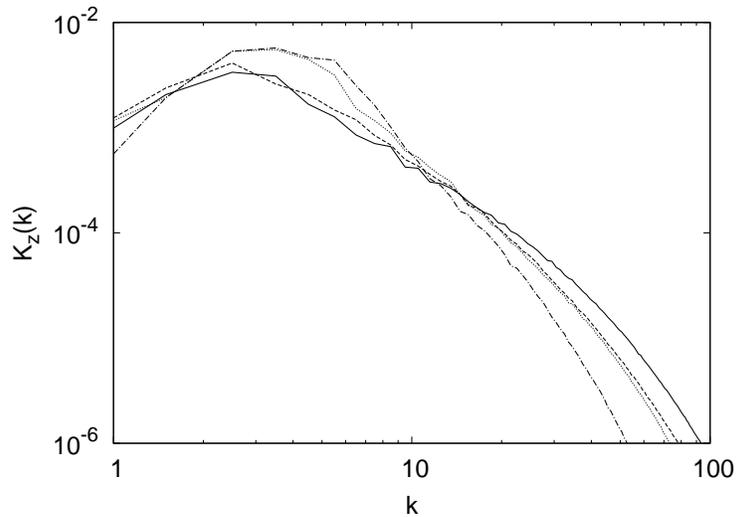}
\caption{
Energy spectra of the vertical velocity component at time $t = 3.1 \tau$
Newtonian flow: solid line.  
Viscoelastic flows: dashed line (Run A), dotted line (Run B), 
dash-dotted line (Run C).
}
\label{fig8}
\end{figure}

The increase of kinetic energy is accompanied by a reduction of viscous
dissipation (Fig.~\ref{fig6}d).  This is a clear fingerprint  of a {\it drag
reduction} phenomenon as defined for homogeneous-isotropic turbulence
\cite{bdgp_pre03,dcbp_jfm05},  i.e. a reduction of turbulent energy 
dissipation at given kinetic energy.
In the present case, a quantitative measure of the drag reduction is 
provided by the ratio between the loss of potential energy and the resulting 
plumes kinetic energy.  
The first can be easily computed by the definition of the potential energy 
$P= -\beta g \langle z T \rangle$ assuming a linear temperature 
profile within the
mixing layer, which gives $\Delta P = P(0)-P(t) \simeq 1/6 Ag h(t)$.  An
estimate of the kinetic energy associated with large scale plumes can be
obtained in terms of the mixing layer growth rate $\dot{h}(t)$ as 
$K_L \sim 1/2 (\dot{h}(t))^2$. We remark that a similar estimation was
proposed by Fermi for modeling the growth of mixing layer (see 
Appendix).  The drag reduction coefficient $f$ is then defined as
\begin{equation} 
f = \frac{\Delta P}{K_L} = 1/3 Ag \frac{h}{\dot{h}^2} = \frac{1}{12 \alpha} 
\label{eq:4}
\end{equation} 
which turns out to be 
inversely proportional to the coefficient $\alpha$ 
which characterizes the mixing layer growth rate \cite{cc_natphys06}.
With this definition, we measure $22\%$ of drag reduction for 
the viscoelastic run B and $30\%$ for the run C (see Fig.~\ref{fig7})

The scenario which emerges from these results is that polymers reduce the
turbulent drag between rising and sinking plumes.  The RT viscoelastic system
is therefore able to convert more efficiently potential energy into kinetic
energy contained in large plumes. Conversely, the turbulent transfer of kinetic
energy to small-scale structures is reduced, which results in a reduction of
the viscous energy dissipation.  This picture is confirmed by the inspection of
the energy spectra (see Fig.~\ref{fig8}). At small scales we found a
suppression of turbulent kinetic energy with respect to the Newtonian case,
while at large scale an increase of the kinetic energy is observed.

The drag reduction scenario therefore provides a clear interpretation
of the effects observed on the mixing properties.
The enhancement of large-scale mixing associated with the faster growth 
of the mixing layer is directly connected to the reduced friction between 
plumes, and the reduced efficiency of small-scale mixing is a natural 
consequence of the suppression of small-scale turbulence.  

\begin{figure}
\includegraphics[clip=true,keepaspectratio,width=10cm]{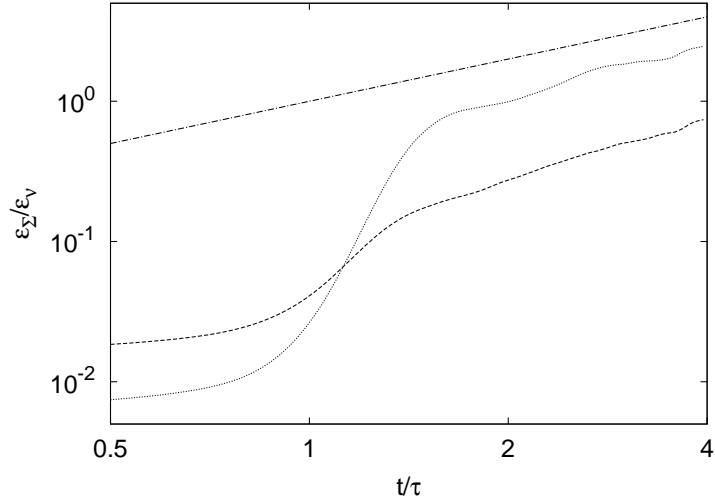}
\caption{Temporal evolution of the ratio between elastic 
and viscous dissipation.  Run A: dashed line. Run B: dotted line.}
\label{fig9}
\end{figure}

The accelerated nature of the RT turbulence poses an interesting question 
about the existence of an asymptotic state for viscoelastic RT.  
For the Newtonian case the phenomenological theory assumes that in the 
late stage of the evolution all terms in the energy balance (\ref{eq:3}) 
have the same temporal scaling determined by gravity forces.  
This implies that 
$-{d P \over dt} \sim \varepsilon_{\nu} \sim t$ and $K \sim t^2$.  
In the viscoelastic case it is not possible to fix a priori the scaling
law for the elastic contribution, because elastic energy
$\Sigma$ is proportional to the elastic dissipation rate 
$\varepsilon_{\Sigma} = 2 \Sigma / \tau_p$.  
Assuming that the latter has the same temporal scaling
than the viscous dissipation,  
$\varepsilon_{\nu} \sim \varepsilon_{\Sigma} \sim t$,
one gets that the elastic contribution to the total energy should become
negligible at long times.  
On the other hand, the assumption that elastic and kinetic
energy have the same scaling $\Sigma \sim K \sim t^2$ leads to the conclusion
that elastic dissipation would eventually dominate over the viscous one. 
Our simulations support the second hypothesis: the ratio between 
elastic end viscous dissipation is not constant, 
and grows almost linearly in time (see Fig.~\ref{fig9}).
A deeper investigation of this asymptotic state in which
polymers are strongly elongated would require to go beyond Oldroyd-B model,
and to adopt more realistic polymer model (e.g. FENE-P model) which 
accounts for maximal elongation and non-linear relaxation.

\section{Heat transport enhancement}
\label{sez:4}
\begin{figure}
\includegraphics[clip=true,keepaspectratio,width=10cm]{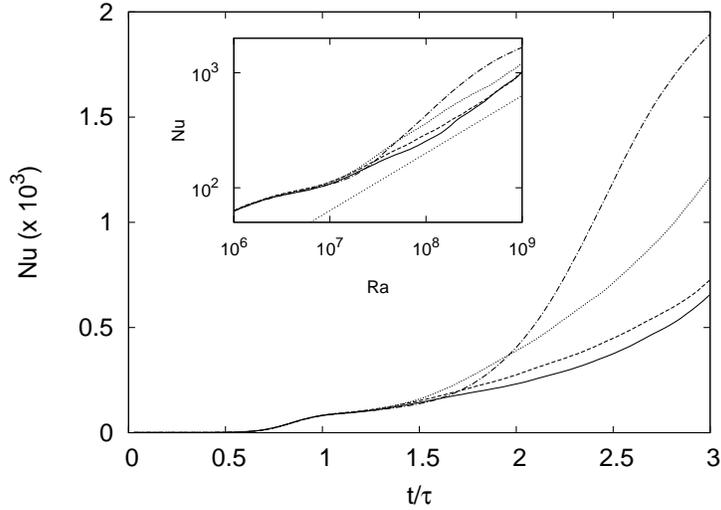}
\caption{
Temporal evolution of Nusselt number 
$Nu=\langle w T \rangle h/(\kappa \theta_0)$.
Inset: Nusselt number vs. Rayleigh number 
$Ra=Agh^3/(\nu \kappa)$. 
Newtonian flow: solid line.  Viscoelastic flows: dashed line (Run A), 
dotted line (Run B), dash-dotted line (Run C).}
\label{fig10}
\end{figure}
\begin{figure}
\includegraphics[clip=true,keepaspectratio,width=10cm]{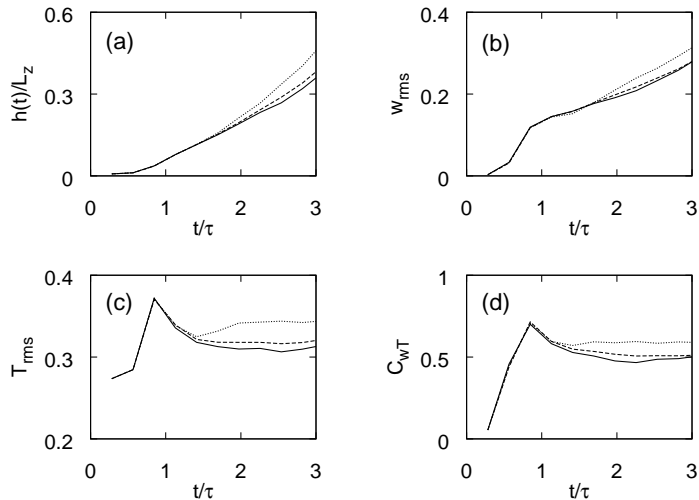}
\caption{Temporal evolution of the contributions to the heat 
transfer efficiency. 
Panel (a): Mixing layer width $h(t)$. 
Panel (b): Rms vertical velocity $w_{rms}$.
Panel (c): Rms temperature $T_{rms}$. 
Panel (d): Correlation between temperature and vertical velocity $C_{wT}(z,t)$. 
Newtonian flow: solid line.  
Viscoelastic flows: dashed line (Run A), dotted line (Run B). 
}
\label{fig11}
\end{figure}

The heat transport efficiency in turbulent convection is usually measured
by the Nusselt number
$Nu=\langle w T \rangle h/(\kappa \theta_0)$, 
which represents the ratio between convective and conductive heat transport.
For a developed turbulent flow the Nusselt
number is expected to behave as a simple power law with respect to  
the dimensionless temperature jump which defines 
the Rayleigh number $Ra=Agh^3/(\nu \kappa)$ \cite{gl_jfm00}. 
For a flow in which boundary layers are irrelevant,
as in our case, Kraichnan \cite{kraichnan62} predicted many years ago 
an asymptotic regime which is expected to emerge at very large $Ra$. 
For this so-called ultimate state regime of thermal convection dimensional
analysis predicts the scaling laws \cite{gl_jfm00}
\begin{equation}
Nu \simeq Pr^{1/2} Ra^{1/2} \qquad
Re \simeq Pr^{-1/2} Ra^{1/2} \, .
\label{eq:5}
\end{equation}
For the case of time dependent RT turbulent convection, all these 
dimensionless quantities depend on time. Dimensionally estimation 
gives (for the Newtonian case)
$Ra \simeq (\beta g \theta_0)^4 t^6/(\nu \kappa)$, 
$Re \simeq (\beta g \theta_0)^2 t^3/\nu$ and
$Nu \simeq (\beta g \theta_0)^2 t^3/\kappa$, which indeed imply
the scaling laws (\ref{eq:5}) and which have been observed 
recently in numerical simulation of RT turbulence 
\cite{bmmv_pre09,bdm_prl10}.

The addition of polymers strongly enhances the efficiency of heat transport, 
i.e. the Nusselt number grows faster both as a function of time and as a 
function of $Ra$  \cite{bmmv_prl10}, and the effects increase with the 
polymer relaxation time, as shown in Fig.~\ref{fig10}.
In order to identify the different causes which contribute to this effect, 
it is useful to rewrite the Nusselt number as: 
\begin{equation} 
Nu = \frac{1}{\kappa \theta_0} h w_{rms} T_{rms} C_{wT}
\label{eq:6}
\end{equation}
where $C_{wT} = \langle wT \rangle/(w_{rms} T_{rms})$ is the correlation
between the vertical velocity component and the temperature field.
In the four panels of
Fig.~\ref{fig11} we plot the four contributions $h$ (panel a) $w_{rms}$ (panel
b), $T_{rms}$ (panel c) and  $C_{wT}$ (panel d) as a function of time.
It is evident that the
increased heat transfer is not simply a consequence of the faster evolution of
the mixing layer $h$, but also of the increased rms of the 
vertical velocity component, $w_{rms}$.
Moreover, the reduction of small-scale turbulent mixing causes 
an increase of the
temperature fluctuations $T_{rms}$ which also gives a positive contribution to
the Nusselt number.  Finally, in the viscoelastic case we found stronger
correlations between temperature and vertical velocity component which
therefore transport heat more efficiently.
In conclusion, the increased heat transport efficiency is a 
combined effect of different contributions: the presence of faster 
thermal plumes, the reduced turbulent mixing, and the stronger 
correlation between thermal plumes and the vertical velocity component. 
While the first contribution is distinctive of RT turbulence, the others
could in principle be observed in other thermal convective systems.
A recent experiment performed within the framework of Rayleigh-Benard 
convection indicates in that case an opposite effect of heat transfer
reduction  \cite{an_prl10} but this can be probably attributed to the 
moderate stretching of polymers in that case.

\section{Conclusions}
\label{sez:5}
The behavior of viscoelastic flows in the RT setup provides the first clear
evidence of simultaneous occurrence of both polymer drag reduction and 
heat transport enhancement.  
Drag reduction in this system is caused by a reduced drag between rising and
sinking thermal plumes, a fact which implies the speed up of the mixing layer
growth. This process shares many analogies with drag reduction observed in
homogeneous, isotropic turbulence, namely the suppression of small-scale
turbulence which results in a reduced viscous drag.  
These analogies provide a support to the conjecture of a common
underlying mechanisms behind these different manifestations of the polymer drag
reduction in bulk flows.

For RT system it is possible to introduce a drag coefficient in terms of
the ratio between the potential energy loss which forces the flow, and the
resulting kinetic energy associated with thermal plumes.  The viscoelastic
case is characterized by faster and more coherent thermal plumes. The effects
on mixing is to enhance the large-scale mixing, and to reduce the small-scale
one. As a consequence, the drag coefficient is reduced and the heat transport 
efficiency, measured by the Nusselt number, is increased. 

We conclude with some speculations on the possible observation of
heat transfer enhancement in laboratory experiments.
The values of the parameters used in our simulations 
can be used to determine the setup for a comparable experiments. 
The units of time $T$ and length $L$ which allow one to convert  
the parameters of our simulations into physical quantities 
can be fixed by matching the numerical values of 
viscosity $\tilde{\nu} = 3 \cdot 10^{-4}$
and gravity $\tilde{g} = (4 A)^{-1}$ used in our simulations 
with physical values $g=9.81 m s^{-2}$, 
$\nu = \nu_{H_2O} = 10^{-6} m^2 s^{-1}$: 
\begin{eqnarray}
L^3 & = & \frac{\tilde{g}}{\tilde{\nu}^2} \frac{\nu^2}{g}\\
T^3 & = & \frac{\tilde{g}^2}{\tilde{\nu}} \frac{\nu}{g^2}
\end{eqnarray}
By choosing the Atwood number $A=0.1$ one gets 
$L \simeq 1.4 cm$ and $T \simeq 0.06 s$. 
This correspond to an experimental box of 
$L_{x,y} \simeq 10 cm$ and $L_z \simeq 20 cm$, 
and polymer relaxation times $\tau_p = 60 ms$ for the case A, 
which is close to realistic relaxation times of long-chain polymers in water. 
The evolution of the system will be quite fast: 
the time required for the mixing layer to invade the 
whole box is estimated to be roughly $2 s$. 
Let us notice that the limit of small Atwood number, required 
in the present study to justify the Boussinesq approximation,
is not a constraint for an experimental setup, where large 
values of $A$ can be obtained by means of some additives 
(e.g. salt) to generate density differences.
It would be interesting to observe experimentally the influence of 
non-Boussinesq effects on the drag reduction phenomenon. 

\acknowledgments
We thank the Cineca Supercomputing Center (Bologna, Italy) for the
allocation of computational resources.

\appendix
\section{Fermi model for the growth of mixing layer}
\label{app1}
Enrico Fermi, together with John von Neumann, were probably the
first who considered a model for the growth of mixing layer 
in the nonlinear stage. The model is described in two reports of 
the Los Alamos Scientific Laboratory, the first from September 1951
(Fermi alone) and the second from August 1953 (Fermi and von Neumann)
never published \cite{fv_lanl55}.
The idea of this work, in the words of the authors, is to 
``discuss in a very simplified form the problem of the growth of
an initial ripple on the surface of an incompressible liquid in 
presence of an acceleration''. The first report of Fermi considers
the interface between a liquid and vacuum, while the second 
report with von Neumann analyzes the case of two fluids of different
densities.

\begin{figure}
\includegraphics[clip=true,keepaspectratio,width=10cm]{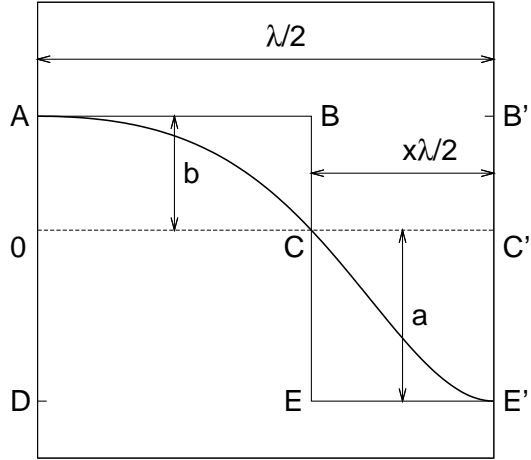}
\caption{Fermi's model for the evolution of the interface. The 
smooth interface $ACE'$ is replaced by the square wave $ABCEE'$.}
\label{fig12}
\end{figure}

The idea of Fermi is to approximate the interface with a square
wave whose shape is characterized by three parameters: the heights
of spike and bubble $a$ and $b$ and the width of the spike $x$
(see Fig.~\ref{fig12}).
Incompressibility gives a relation among these quantities,
$b=a x/(1-x)$. Fermi next considers the Euler-Lagrange equations
for the variation of the potential and kinetic energy and 
obtains a couple of equations for the evolution of $a$ and $x$.
In the following we consider a simplified version of Fermi model
with bubble-spike symmetry ($b=a$, $x=1/2$), consistent with the Boussinesq 
approximation discussed in the present paper.

The variation of potential energy to generate the profile in 
Fig.~\ref{fig12} is 
\begin{equation}
U = {\rho_2 - \rho_1 \over 2} g L_x L_y a^2
\label{eq:app1}
\end{equation}
For the kinetic energy, assuming that the ``plumes'' $ABCO$ and $CC'E'E$
move respectively up and down with velocity $\dot{a}$ and plumes 
$BB'C'C$ and $OCED$ move respectively right and left with the same
velocity (for incompressibility) one obtains
\begin{equation}
K = {\rho_1 + \rho_2 \over 2} L_x L_y a \dot{a}^2
\label{eq:app2}
\end{equation}
From the Lagrange equations
\begin{equation}
{d \over dt} {\partial K \over \partial \dot{a}} - 
{\partial K \over \partial a} = - {\partial U \over \partial a}
\label{eq:app3}
\end{equation}
one the obtains a differential equation for the $a(t)$ without 
free parameter. We remark that (\ref{eq:app3}) assumes that all 
potential energy is transformed in large scale kinetic energy 
generating the motion of the interface. In a later stage, in which
a turbulent flow develops, we can still try to use (\ref{eq:app3})
but now with a factor $0 \le \delta \le 1$ in front of the rhs
which takes into account that a fraction ($1-\delta$)
of potential energy goes in viscous dissipation (through 
the turbulent cascade). With this 
correction the equation for $a(t)$ reads
\begin{equation}
{d^2 a \over dt^2} a + {1 \over 2} \left( {da \over dt} \right)^2=
{1 \over 4} \delta A g a
\label{eq:app4}
\end{equation}
Introducing the width of the interface $h(t)=2 a(t)$ and solving
(\ref{eq:app4}) with initial condition $h(0)=h_0$ and replacing 
$\delta \equiv 8 \alpha$ we finally get
\begin{equation}
h(t) = h_0 + (4 \alpha A g h_0)^{1/2} t + \alpha A g t^2
\label{eq:app5}
\end{equation}
which is the form proposed from other authors on the basis of 
completely different considerations \cite{rc_jfm04,cc_natphys06}.


\bibliography{biblio2}

\end{document}